\documentstyle[11pt,epsfig,amssymb,citesort]{article}
\newcommand{\beq}{\begin{eqnarray}}
\newcommand{\eeq}{\end{eqnarray}}

\newcommand{\np}{Nucl. Phys.\ }

\newcommand{\asgen}{\alpha_s}

\newcommand{\as}{\alpha_s{{}^{\rm MOM}}}
\newcommand{\astil}{\alpha_s{{}^{\widetilde{\rm MOM}}}}

\newcommand{\Lams}{\Lambda_{\overline{\rm MS}}}

\newcommand{\be}{\begin{equation}}
\newcommand{\ee}{\end{equation}}
\newcommand{\lwrsim}{\raise0.3ex\hbox{$<$\kern-0.75em\raise-1.1ex\hbox{$\sim$}}}
\def\Am#1#2#3{\widetilde A_{#1}^{#2}(#3)}
\def\A#1#2#3{A_{#1}^{#2}(#3)}
\def\C2#1#2{({\cal C}_2)_{#1}^{#2}}

\def\glandauM#1#2#3{(g_{#1 #2}-\frac{#3_#1 #3_#2}{-k^2})}

\def\jhep#1#2#3{J. High Energy Phys. {\bf #1} (#2) #3}
\def\prd#1#2#3{Phys.\ Rev.\ {\bf D#1} (#2) #3}
\def\npb#1#2#3{Nucl.\ Phys.\ {\bf B#1} (#2) #3}
\def\plb#1#2#3{Phys.\ Lett.\ {\bf B#1} (#2) #3}

\def\np#1#2#3{Nucl.\ Phys.\ B#1 (19#3) #2}

\begin{document}
\setcounter{page}{1}
\begin{flushright}
LPT-ORSAY 00/70\\
UHU-FT/00-02\\
\end{flushright}
\begin{center}
\bf{\huge 
Consistent OPE Description of Gluon Two- and Three-point Green Functions ?}
\end{center}  
\vskip 0.8cm
\begin{center}{\bf  Ph. Boucaud$^a$, A. Le Yaouanc$^a$, J.P. Leroy$^a$, 
J. Micheli$^a$, \\ 
O. P\`ene$^a$, J. Rodr\'\i guez--Quintero$^b$   
}\\
\vskip 0.5cm 
$^{a}$ {\sl Laboratoire de Physique Th\'eorique~\footnote{Unit\'e Mixte 
de Recherche du CNRS - UMR 8627}\\
Universit\'e de Paris XI, B\^atiment 210, 91405 Orsay Cedex,
France}\\
$^b${\sl Dpto. de F\'{\i}sica Aplicada e Ingenier\'{\i}a el\'ectrica \\
E.P.S. La R\'abida, Universidad de Huelva, 21819 Palos de la fra., Spain} \\
\end{center}
\begin{abstract}

We perform an OPE analysis of the flavorless non-perturbative  gluon propagator
and the symmetric three-gluon vertex in the Landau gauge.  The first subdominant
operator is $A_\mu A^\mu$ which can condensate in the  Landau gauge ``vacuum''
although being a non-gauge invariant operator.   We neglect all higher dimension
operators. Then the gluon propagator and the symmetric  three gluon vertex only
depend on one common  unknown condensate. We propose a consistency check  from
lattice data. At two loops for the leading coefficient and with $1/p^2$
corrections at tree-level order the two fitted values for the condensate do not
agree. At three loops we argue that the today unknown $\beta_2^{\rm MOM}$ should
be equal to  $1.5(3)\times \beta_2^{\widetilde{\rm MOM}}=7400(1500)$ to fulfill the OPE
relation. Inclusion of the  power corrections' anomalous dimensions  should
improve further the agreement. We show that these techniques cannot be applied to
the asymmetric three gluon vertex with one vanishing momentum.

\medskip

\noindent P.A.C.S.: 12.38.Aw; 12.38.Gc; 12.38.Cy; 11.15.H

\end{abstract}

\section{Introduction}
\label{sec:intro}
 
Much effort has being devoted in the last years to the non-perturbative 
computation of the running QCD coupling constant $\asgen(p)$
\cite{Lusch,Bali,otros,cpcp,frenchalpha,propag,poweral}. A most surprising result  
arising from  the lattice
computation of the gluon three-point Green function \cite{poweral} is that a
$\sim 1/p^2$  power correction  is still sizeable for momenta  of the order of
$\sim 10$ GeV !!

These  power corrections   were qualitatively discussed in \cite{poweral} in
connection with several physics phenomena such as interquark potential,  the
gluon condensate puzzle, Landau pole, etc... It was also strongly stressed that
$\asgen(p)$ being non-perturbatively defined on a lattice in the {\it particular}
Landau gauge, a contribution to the operator product expansion (OPE) from a gauge
dependent  local operator {\it should not a priori be} neglected. More precisely,
{\it the $A^2=A_\mu A^\mu$ operator can  acquire a vacuum expectation value
(v.e.v.) because of the non-gauge-invariant nature  of the vacuum}. As a matter
of fact, the authors of ref. \cite{lavelle} showed in a general way that the OPE
for QCD propagators unequivocally generates non null {\it v.e.v.} of $A_\mu
A^\mu$. The important point is that in contrast to that of the gauge invariant
$G^{\mu \nu} G_{\mu \nu}$ generating $1/p^4$ powers, it produces $1/p^2$
corrections which may be  of comparable size to higher perturbative correction
(three-loops, for instance) for the considered momenta. 

The key remark here is that $A^2$ is the only dimension-two operator able to
generate a non vanishing {\it v.e.v.} The purpose of the present paper is to
exploit this fact according to OPE philosophy in the following manner: we will
study both the gluon propagator and the three gluon vertex (two- and three-point
Green functions) in order {\it to examine whether the same $\langle A^2 \rangle$
condensate can consistently describe the evaluated  power corrections to both
quantities}. For rather obvious reasons which will be explicited later, we will
totally neglect $O(1/p^4)$ corrections.

The OPE approach consists in expanding the Green functions in inverse power
series in $p^2$, every term corresponding to a local operator. The  coefficients
which multiply the operators are calculable in perturbation\cite{weinberg}. In
practice of course the coefficients are expanded to a given finite order, not
only because the perturbative series are only asymptotic, but also because
computing  higher orders is a rapidly increasing task which has not yet been
performed.

The potential dangers when using OPE to relate different Green functions via the
same condensates has been stressed by several authors  \cite{david,MartiSach}: it
was claimed that the coefficients of the leading operator at any given momentum
scale have to be computed to a sufficiently  high perturbative order to make the
perturbative uncertainty smaller than the power corrections.

Our input data are lattice computed Green functions 
\cite{frenchalpha,propag,poweral}, covering a momentum  range from 2
to 10 GeV. We will analyze them according to OPE.  The gluon propagator and the
symmetric three-point Green function ($p_1^2=p_2^2=p_3^2$) can be expanded into
perturbatively computable coefficients and the only unknown $\langle A^2 \rangle$
condensate.  This allows to perform a test of OPE in a relatively favorable
situation. Our first ``{\it ace}'' is indeed this wide momentum window which 
reduces the risk of confusion between neglected perturbative (logarithmic)
contributions and the $\sim 1/p^2$ terms\footnote{Still it should be noted that
the recent perturbative computation of  $\beta_3$ in asymmetric MOM
scheme\cite{Chetyrkin} leads to a four-loop estimate of the  empiric power
coefficient introduced in ref. \cite{poweral} wich is roughly $0.8$ times the
three-loop one. This shows that some amount of disguise of higher order
perturbative terms into power corrections is difficult to fully avoid.}. Our
second  ``{\it ace}'' is that  $1/p^2$ terms are rather easy to display in the 2
- 10 GeV window \cite{poweral} while, for example, $1/p^4$ contributions
would only influence the lower part of this window making the fits hazardous.
Our third   ``{\it ace}'' is the high accuracy achieved by our lattice
calculations.

The leading coefficient (the coefficient of identity operator in the OPE) of the
gluon propagator  is known up to three loops but the symmetric three-gluon 
vertex is only known at two loops. To ask for an OPE consistent description when
leading power coefficients are  only expanded at two loops might be exposed to
the danger mentioned above \cite{MartiSach}. We intend to check this.  We will
also try  the three-loop consistent description by fitting the unknown
perturbative information for  the three-gluon vertex, the $\beta_2$ parameter of
beta function in MOM scheme.  The comparison with future perturbative
computations will obviously be in order. 

We have only computed  at tree-level order the perturbative  coefficients
multiplying the   $A^2$ operator. We will therefore perform the fits  with a
constant times $1/p^2$ term. A knowledge of the anomalous dimension of the
coefficient which multiplies $1/p^2$ will be welcome \cite{next}.

The theoretical setting of our use of OPE is described in Section 
\protect\ref{sec:model}. A preliminary analysis of results from
previous lattice data are presented in Section 
\protect\ref{sec:lattice}. We finally discuss and conclude in Section 
\protect\ref{sec:conclusions}.

\section{OPE for the gluon propagator and  $\asgen(p)$}
\protect\label{sec:model}

The purpose of the present section is to develop an OPE-based model   describing
the non-perturbative power behaviour of gluon propagator and $\asgen(p)$. To this
goal, we will compute the first power correction in  OPE (always {\it at
tree-level}) for two and three gluon operators, by following the prescriptions of
\cite{weinberg}. Some general results for the gluon propagator in full QCD can be
found in \cite{lavelle}. For completeness we will  describe our own calculations
for both the propagator and the three gluon vertex. We have checked agreement for
the propagator with \cite{lavelle}.

 The two- and three-point bare lattice Green functions are renormalized by
imposing momentum subtraction prescriptions. Let us insist:  such a
renormalization prescriptions leads to the definition of a gauge-dependent
coupling which consequently admits  a non vanishing contribution from the
non-gauge invariant OPE $A_\mu A^\mu$.

We work in the pure Yang-Mills QCD, without quarks. OPE yields  

\beq
T\left( \Am{\mu}{a}{-p} \Am{\nu}{b}{p}\right)&=& \nonumber \\
(c_0)^{a b}_{\mu \nu}(p) \ 1 
&+& (c_1)^{a b \mu'}_{\mu \nu a'}(p) : \A{\mu'}{a'}{0}: \ + \
(c_2)^{a b \mu' \nu'}_{\mu \nu a' b'}(p) \ 
:\A{\mu'}{a'}{0} \ \A{\nu'}{b'}{0}:  \nonumber \\
&+&\dots \;\; , \label{OPEfield1} 
\eeq
\beq
T\left( \Am{\mu}{a}{p_1} \Am{\nu}{b}{p_2} \Am{\rho}{c}{p_3} \right)&=& \nonumber \\
(d_0)^{a b c}_{\mu \nu \rho}(p_1,p_2,p_3) \ 1 
&+& (d_1)^{a b c \mu'}_{\mu \nu \rho a'}(p_1,p_2,p_3) : \A{\mu'}{a'}{0}:
\nonumber \\ 
&+& (d_2)^{a b c \mu' \nu'}_{\mu \nu \rho a' b'}(p_1,p_2,p_3) \   
:\A{\mu'}{a'}{0} \ \A{\nu'}{b'}{0}: \  + \
\dots \;\; ; \nonumber \\
\label{OPEfield2}
\eeq

\noindent written only in terms of normal products of local gluon  field
operators, where $A$ (${\widetilde A}$) stands for the gluon field in 
configuration (momentum) space, $a,b$ being colour indices and $\mu,\nu$ Lorentz
ones. The notation $T()$ simply refers to the standard $T^\ast$ product  in
momentum space. The normal product of Eqs. (\ref{OPEfield1},\ref{OPEfield2}) will
be defined below in reference to the perturbative vacuum. The  r.h.s. in eqs.
(\ref{OPEfield1},\ref{OPEfield2}) contains operators of dimension up to two. We
have omitted local operators including,  for instance, gluon field derivatives or
ghosts. Indeed  at the order $1/p^2$,  which we consider in this letter, they
have a vanishing {\it v.e.v.}  (Obviously $<\partial_\mu A^\mu>=0$). 

\subsection{Gluon propagator} 

The following step is to take the QCD {\it v.e.v.} (in our 
flavorless universe) in both r.h.s. and l.h.s. of equation (\ref{OPEfield1}),

\beq
\langle T\left(\Am{\mu}{a}{-p} \Am{\nu}{b}{p}\right) \rangle  \ = \
(c_0)^{a b}_{\mu \nu}(p^2) \ + \ (c_2)^{a b \mu' \nu'}_{\mu \nu a' b'} \
\delta^{a' b'} g_{\mu' \nu'} \ \frac{\langle A^2 \rangle}{4 (N_c^2-1)} 
+ \dots \; \; ; 
\label{OPEpropag}
\eeq

\noindent where $A^2= \ :\A{\mu'}{a'}{0} \ \A{a'}{\mu'}{0}: \ $. A first
immediate simplification in r.h.s. of Eq. (\ref{OPEpropag}) with regard to Eq. 
(\ref{OPEfield1}) is the disappearance of terms containing an odd  number of
local gluon field operators. They contribute only at higher orders in the OPE
because of the impossibility to build a Lorentz invariant tensor (with odd number
of indices) without any external non-null momentum.  The normal product
in eqs. (\ref{OPEfield1},\ref{OPEfield2}) refers to the perturbative vacuum
$|0>$. To be more specific the non-perturbative condensate
$<A^2>$ is defined  from eq. (\ref{OPEpropag}) as follows
\footnote{We thank Y. Dokshitzer and G. Korchemsky for illuminating discussions
on this point.} :  given the l.h.s computed non-perturbatively on a
lattice\footnote{Being a gauge dependent quantity it cannot be accessed from
experiment.}, given the coefficients $c_0$ and $c_2$ in the r.h.s computed
perturbatively at a given order in perturbation theory and for a particular
renormalization momentum scale and scheme $<A^2>$ is unambiguously defined.

The standard method  \cite{weinberg}  to compute the perturbative expansion 
of the OPE Wilson coefficient  is to compute to the wanted order an appropriate 
matrix element of the l.h.s. operator of Eqs. (\ref{OPEfield1},\ref{OPEfield2}). The 
perturbative gluon propagator 
\beq
(G^{(2)}_{pert})^{a b}_{\mu \nu}(p^2)=\langle 0 | T\left(\Am{\mu}{a}{-p}
\Am{\nu}{b}{p}\right)  | 0 \rangle
\eeq
at the considered order in  perturbation
theory yields $c_0$ while $c_2$ may be obtained  from  
\beq
\langle g |
T\left(\Am{\mu}{a}{-p} \Am{\nu}{b}{p}\right)  | g \rangle_{\rm connected}
=  (c_2)^{a b \mu' \nu'}_{\mu \nu a' b'}(p) \ 
 \langle g\ |:\A{\mu'}{a'}{0} \ \A{\nu'}{b'}{0}:| g \rangle\label{gadeuxg}
\eeq
 with two soft external gluons. From 
\beq 
 \langle g| A^2 |g\rangle = 2 + O(\alpha),
\label{gadeuxg2} \eeq
we may compute $c_2$. At tree-level order
the $O(\alpha)$ is neglected and eq. (\ref{gadeuxg}) allowing to compute
$c_2$ from the computation of the l.h.s. :
\footnote{Beyond  tree-level order the radiative corrections,
$O(\alpha)$ terms in (\ref{gadeuxg2}),  
are calculable, but we will skip this issue in this paper.}

\beq
(c_2)_{\mu \nu a' b'}^{a b \mu' \nu'}=\frac{1}{2}\
\frac{\langle \Am{\tau}{t}{0} \Am{\mu}{a}{-p} \Am{\nu}{b}{p} \Am{\sigma}{s}{0}
 \rangle}
{(G^{(2)}_{pert}(0))^{t a'}_{\tau \mu'} \ 
(G^{(2)}_{pert}(0))^{s b'}_{\sigma \nu'}}.
\label{Diag}
\eeq

\begin{figure}
\hspace*{-1.3cm}
\begin{center}
\begin{tabular}{ll}
\epsfxsize5.5cm\epsffile{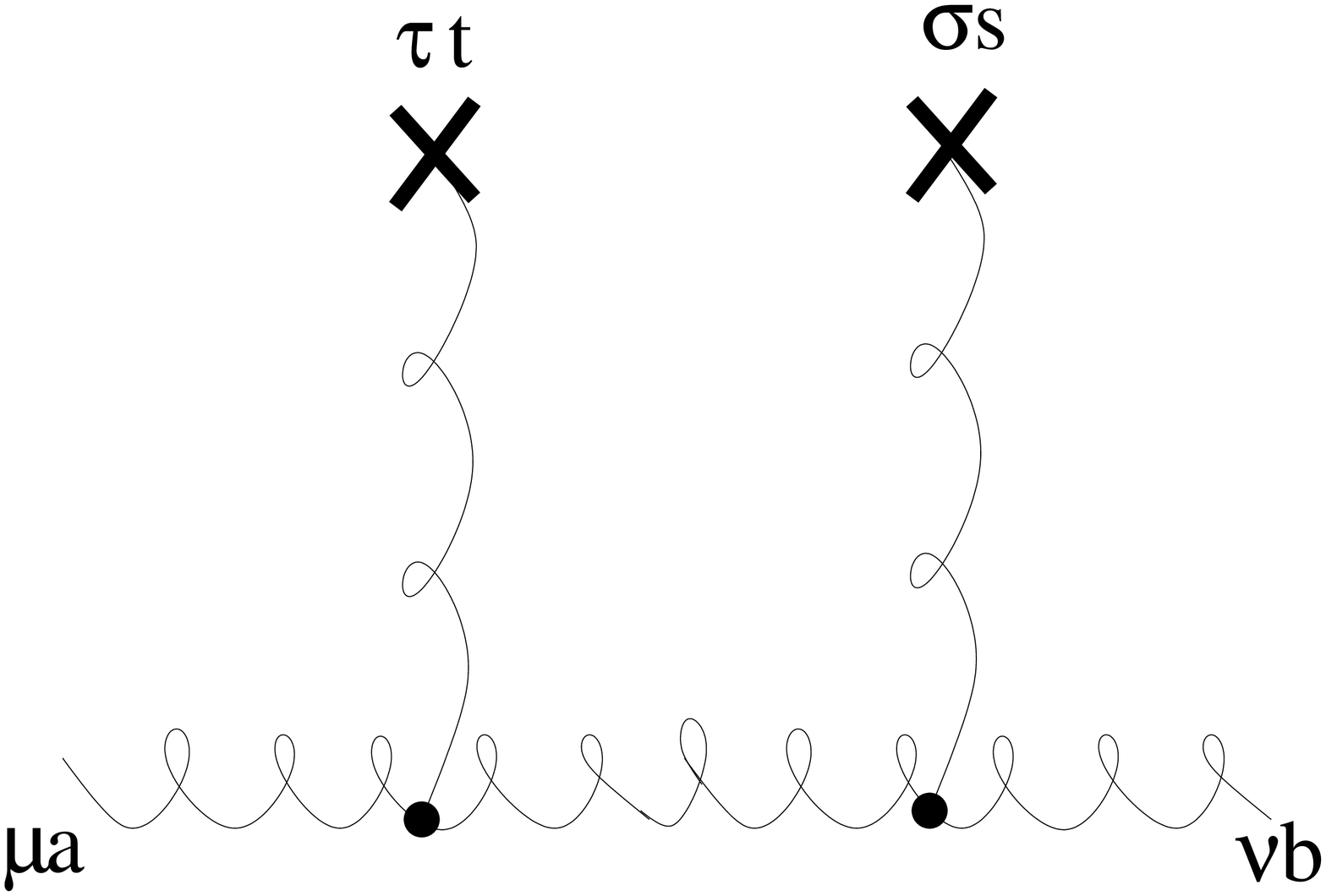} & 
\epsfxsize5.5cm\epsffile{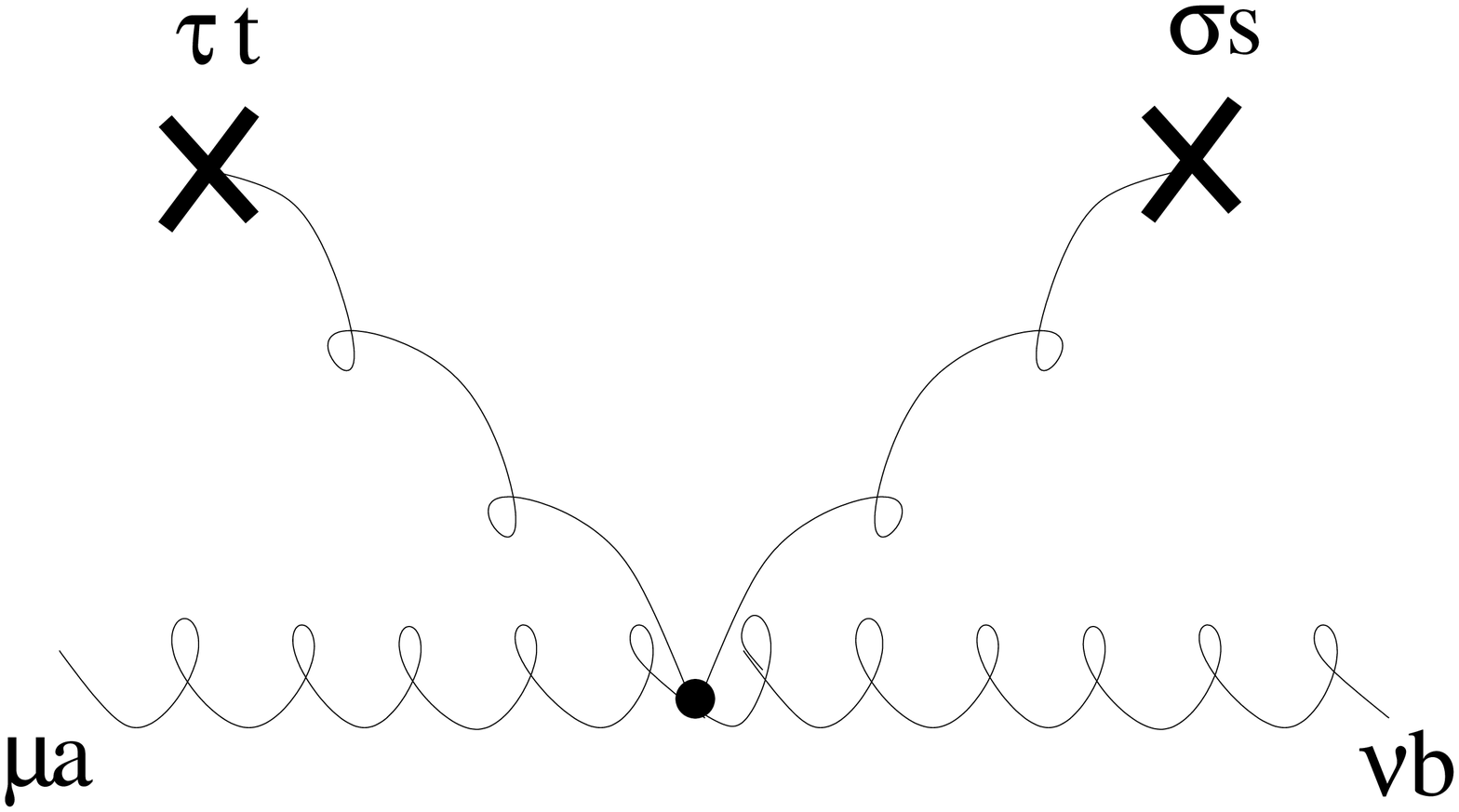} \\
\end{tabular}
\caption{\small {\it Four gluons tree-level diagrams contributing (with all their 
possible permutations) to the  $(c_2)_{\mu
\nu t s}^{a b \tau \sigma}$ OPE perturbative coefficient of the gluon 
propagator. Crosses mark the gluon legs due to the external
soft gluons.}}
\label{Fig1}
\end{center}
\end{figure}

\noindent The ratio in the r.h.s of Eq. (\ref{Diag}) represents symbolically
diagrams with four gluon legs where the two of them carrying zero momentum  are
explicitly {\it cut}.  Consistently with Eq. (\ref{gadeuxg2}), we compute in Landau gauge
the {\it tree-level} diagrams shown in fig. \ref{Fig1} to obtain  $\Gamma_{\mu
\nu t' s'}^{a b \tau' \sigma'}$ and then, for $p^2=-k^2$, we can write:

\beq   
\langle T\left( \Am{\mu}{a}{-p} \ \Am{\nu}{b}{p} \right) \rangle \ = \
\frac{-i}{-k^2} \glandauM{\mu}{\nu}{p} \ \delta^{a b} \nonumber \\ 
\times \left( -k^2 G^{(2)}_{(\rm n \ loops)}(k^2) \
+ \ \frac{3g^2 \langle A^2 \rangle}{4(N_c^2-1)} \ \frac{1}{-k^2} \
+ O\left(\alpha^{n+1},\alpha^2 \frac{\Lambda^2}{-k^2}\right) \right) \; ,
\label{OPEpropagfin}
\eeq

As already mentioned, there is a series of problems in the practical use  of OPE
tried in this paper \cite{david},\cite{MartiSach}.  In general,  the whole 
expansion in terms of condensates suffers from the so-called renormalon
ambiguities : the resummation of the non-Borel-summable perturbative series for
the leading operator Wilson coefficient is ambiguous, this ambiguity being
compensated by  the ultraviolet renormalon ambiguity of the matrix element of the
subdominant operator $A^2$.  By luck the infrared renormalon ambiguity cancels
off when comparing  the contribution of the condensate $<A^2>$  in both the
propagator and the three gluon vertex. In this way we may simply decide to
truncate the perturbative series to some order and the universality of the
condensates is  preserved provided this order is clearly stated. 

A remnant of this renormalon problem arises from the truncation of  OPE  and
perturbative  series for the Wilson coefficients.  One has to be sure to have gone
far enough in perturbative expansions and in the inverse power expansion to have
a coherent scheme. We will try to justify later our truncation of the power
series to order $1/k^2$. For the perturbative series of the leading contribution 
we have to choose a truncation order $n$  according to a compromise: it should be
high enough  to make sure of the dominance of non-perturbative  contributions
compared to the perturbative uncertainty from neglected higher orders;  but low
enough not to destroy the good asymptotic behaviour of the series. 

The gluon propagator turns out to be known up to the third  loop in its perturbative expansion
\cite{Davy,propag}, but not yet the  symmetric MOM coupling constant. We can thus try the exercise
with both $n=2$ and  $n=3$, but making in the latter case some fit for the unknown three loop
coefficient  ($\beta_2$). This is not too high an order\footnote{In the $\widetilde{\rm MOM}$
scheme the fourth loop coefficient of the $\beta$ function is known\cite{Chetyrkin} and the
series looks like being in its perturbative regime for momenta higher than $\sim 3$ GeV} over
the considered momentum window but it might be too low. Concerning power corrections higher than
$1/k^2$: our fits do not show the necessity for $1/k^4$ terms (which would exhibit a very
distinguishable functional dependence) over  the large momenta window we consider. This is lucky
since in practice it would be hopeless to deal with $1/k^4$ terms, too many operators being
needed as well as, for consistency, many higher orders in the perturbative expansion of the
$c_0, d_0, c_2$ and $d_2$ coefficients. Anyhow the consistency of the matching to lattice
non-perturbative data for the two- and three-point Green functions will allow to test the
validity of our assumptions.

In brief,  within the scope of the present paper 
we will satisfy ourselves by modeling power contributions to our Euclidean 
two-point Green function,

\beq
\langle \Am{\mu}{a}{-p} \ \Am{\nu}{b}{p} \rangle \ = \ 
\delta_{a b}\left(\delta_{\mu \nu}-\frac{p_\mu p_\nu}{p^2}\right) \ G^{(2)}(p^2) \; ,
\label{Eucprop}
\eeq

\noindent such that

\beq
k^2 G^{(2)}(k^2) \ = \ Z^{MOM}(k^2) \ = \ Z^{MOM}_{\rm n loops}(k^2) \ + 
\frac{3g^2 \langle A^2 \rangle}{4(N_c^2-1)} \ \frac{1}{k^2}  \; ,
\label{OPEescalar2p}
\eeq

\noindent where $Z^{MOM}_{\rm n loops}= k^2 G^{(2)}_{\rm n loops}$, $n=2,3$ and
by studying  systematically the stability of the matching  of lattice  results to
formula (\ref{OPEescalar2p}) for different choices of the momentum window. 

All the quantities in Eqs (\ref{OPEpropag}-\ref{OPEescalar2p}) are bare objects,
where  the explicit dependence on the regularization cut-off has been omitted
for the sake of simplicity. However,  applying MOM prescription at a certain
renormalization scale $\mu^2$, a renormalized version of  Eq.
(\ref{OPEescalar2p}) is obtained, 

\beq
k^2 G^{(2)}_R(k^2,\mu^2) \ = \ Z^{MOM}_{\rm R,n loops}(k^2,\mu^2) \ + 
c(k^2,\mu^2) \ \frac{\langle A^2 \rangle_{R,\mu^2}}{4(N_c^2-1)} \; ,
\label{OPEescalarR2p}
\eeq

\noindent where, in MOM philosophy, $\langle A^2 \rangle_{R,\mu^2}$ is
renormalized  at $\mu^2$ such that the Wilson coefficient takes
the tree-level value computed from the diagrams of fig. \ref{Fig1} : 
$c(\mu^2,\mu^2)= 3(g_R(\mu^2))^2/\mu^2$.  $Z^{MOM}_{\rm R,n loops}(k^2,\mu^2)$ in
 Eq. (\ref{OPEescalarR2p})  behaves according
to the three-loop (two-loop) anomalous dimension of the gluon propagator
\cite{propag} and is such that $\mu^2 G^{(2)}_R(\mu^2,\mu^2)=1$. This last
condition  imposes an overall normalization which is totally irrelevant for our
lattice data  fitting strategy. In this respect the situation
will slightly differ in the next section. By definition the running $k^2$ 
is there the renormalization point for the running coupling constant.
It is then the condensate renormalized
at $k^2$, $\langle A^2 \rangle_{R,k^2}$, and the coefficient accordingly
renormalised at $k^2$ which have to be dealt with. 
This has to be taken into account when computing the anomalous dimension
of the factor multiplying $1/k^2$ in each case as will be done soon 
 \cite{next}. Nevertheless, the
goal of the present paper is to develop the principle of an OPE testing procedure
and to perform a preliminary check. In the scope of this preliminary check, 
where only tree-level power corrections are taken into account, we will simply
take zero anomalous dimensions :

\beq
c(k^2,\mu^2)\langle A^2 \rangle_{R,\mu^2}=3g_R^2 \ \frac{\langle A^2 \rangle_R}{k^2} \ ,
\label{ctree}
\eeq

\noindent where $g_R^2 \ \langle A^2 \rangle_R$ is taken to be a non-perturbative constant
which can be fitted from lattice data.

\subsection{Running coupling constant} 

In momentum subtraction schemes, in Landau gauge, the renormalized coupling
constant can  be defined at the renormalization momentum scale $k^2$
as (See for instance \cite{frenchalpha}) :

\beq
g(k^2)=\frac{G^{(3)}(p_1^2,p_2^2,p_3^2) (Z^{MOM}(k^2))^{3/2}}
{G^{(2)}(p_1^2) G^{(2)}(p_2^2) G^{(2)}(p_3^2)} \ ,
\label{gdef}
\eeq

\noindent where $Z^{MOM}$ and $G^{(2)}$ were defined in the previous subsection,
and  $G^{(3)}(p_1^2,p_2^2,p_3^2)$ is the three-point scalar form factor. Eq.
(\ref{gdef}) is directly written for an Euclidean metric. Nevertheless, as done
in the previous section,  we prefer to compute the scalar form factor in the more
familiar Minkowskian metrics and  translate the final result to Euclidean. This
form factor should be obtained by   projecting of the three-point Green
function,  $\langle \Am{\mu}{a}{p_1} \Am{\nu}{b}{p_2} \Am{\rho}{c}{p_3}\rangle$,
over the  transverse tree-level three-gluon tensor,

\beq
(T^{\rm tree})_{\mu_1 \mu_2 \mu_3}=\left[ g^{\mu'_1 \mu'_2} (p_1-p_2)^{\mu'_3} 
+ \rm{cycl. perm.} \right] \ \prod_{i=1,3} \left( g_{\mu'_i \mu_i}-
\frac{p_{i \mu'_i} p_{i \mu_i}}{p_i^2} \right) , \nonumber \\
\label{Ttree}
\eeq

\noindent considered for a given configuration of the three momenta with an 
overall renormalization scale $k^2$. In the following we shall consider the two 
standard momentum subtraction kinematics: 
$p_1^2=p_2^2=p_3^2=-k^2$ (MOM) and $p_1^2=p_3^2=-k^2$, 
$p_2^2=0$ ($\widetilde{\rm MOM}$). 

In the MOM case, the scalar form factor, $G^{(3)}(k^2,k^2,k^2)$, is obtained from:

\beq
G^{(3)}(k^2,k^2,k^2) \ f^{a b c} \ = \ \frac{1}{-18k^2}
\langle \Am{\mu}{a}{p_1} \Am{\nu}{b}{p_2} \Am{\rho}{c}{p_3}\rangle \nonumber \\
\times \left[(T^{\rm tree})^{\mu_1 \mu_2 \mu_3}+\frac{(p_1-p_2)^\rho (p_2-p_3)^\mu 
(p_3-p_1)^\nu}{-2k^2} \right] \; .
\eeq

\begin{figure}[hbt]
\begin{center}
\begin{tabular}{ccc}
\epsfxsize4.0cm\epsffile{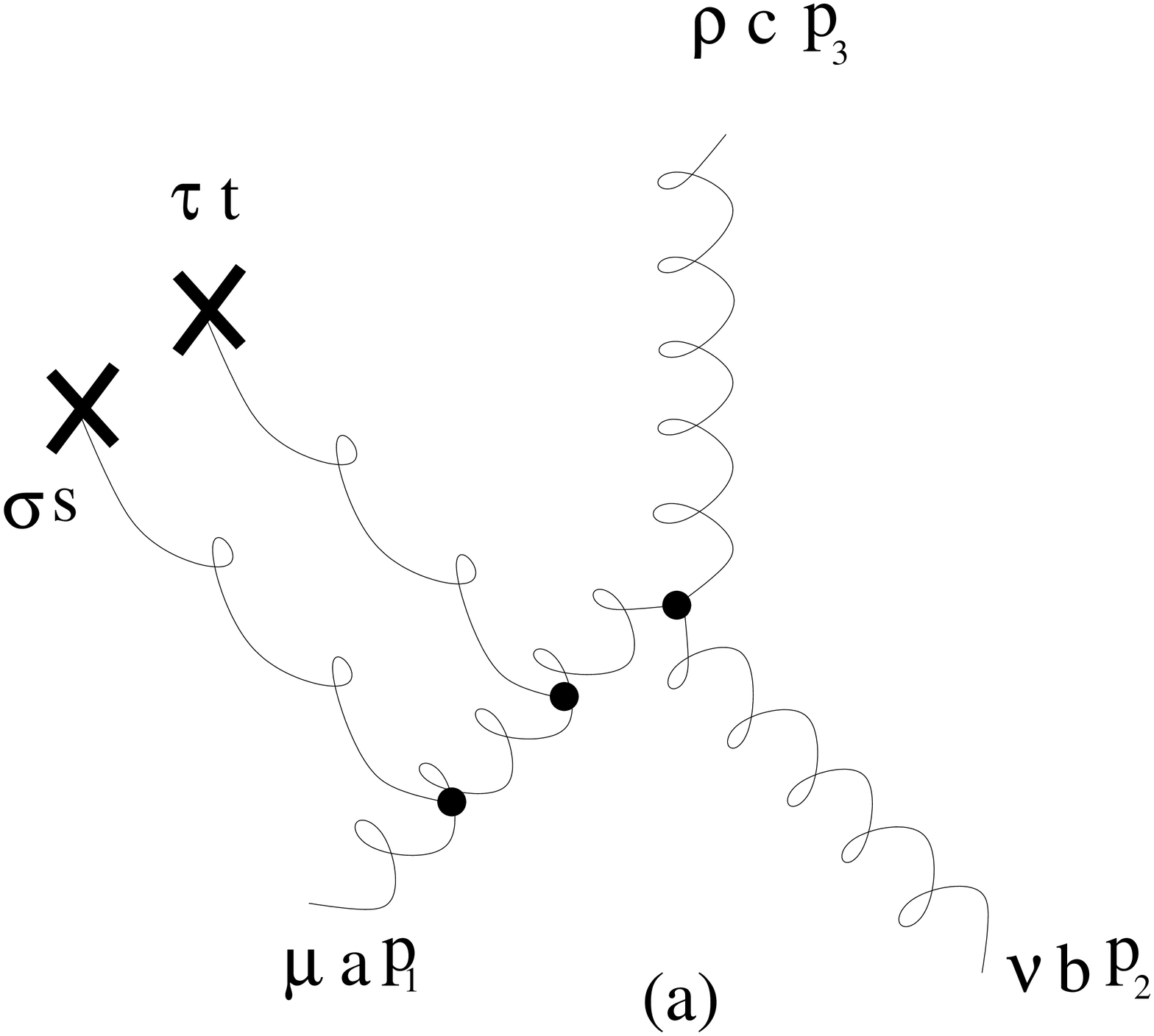} & \hspace{2.5cm} &
\epsfxsize4.0cm\epsffile{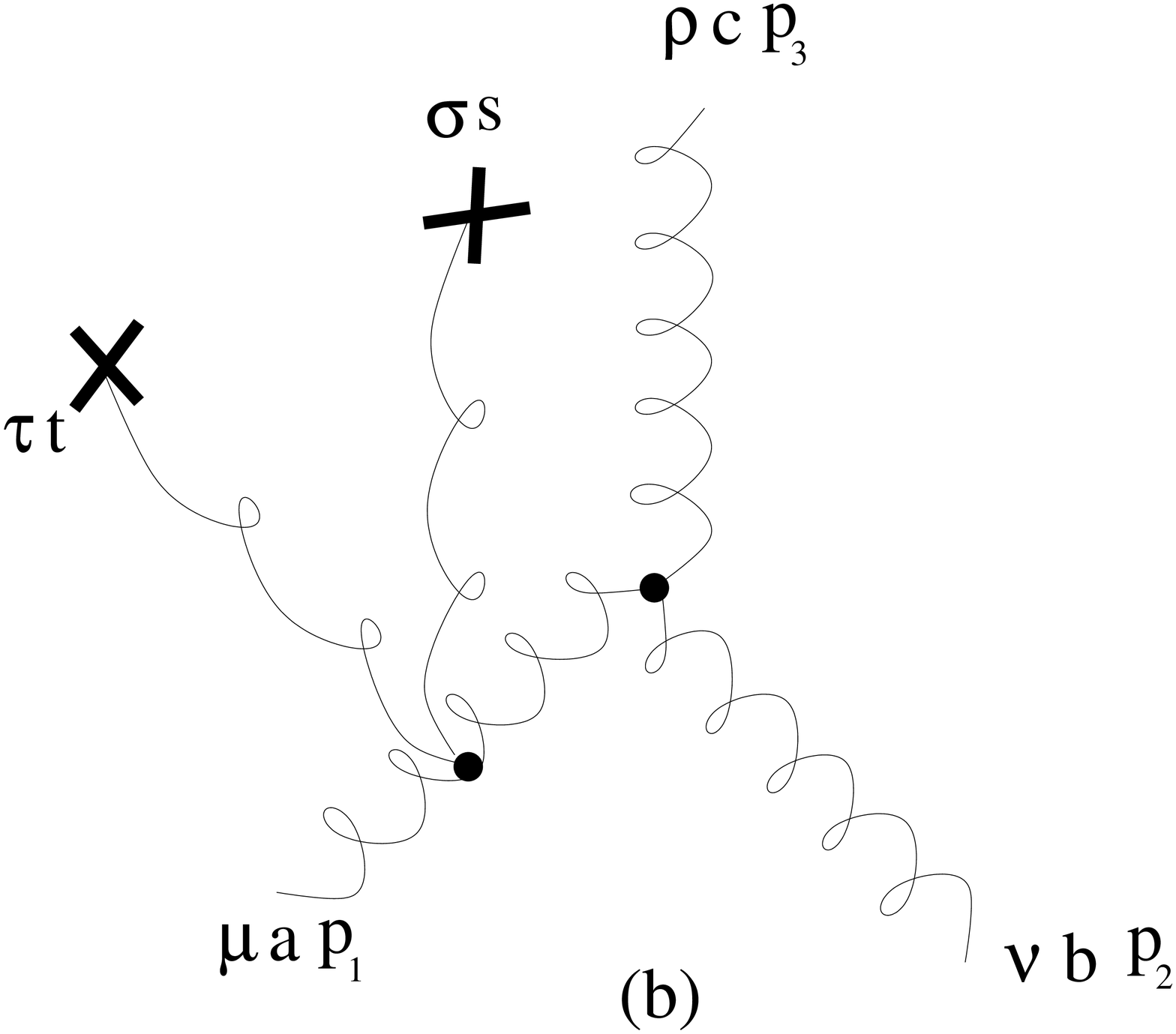} \\
\epsfxsize3.5cm\epsffile{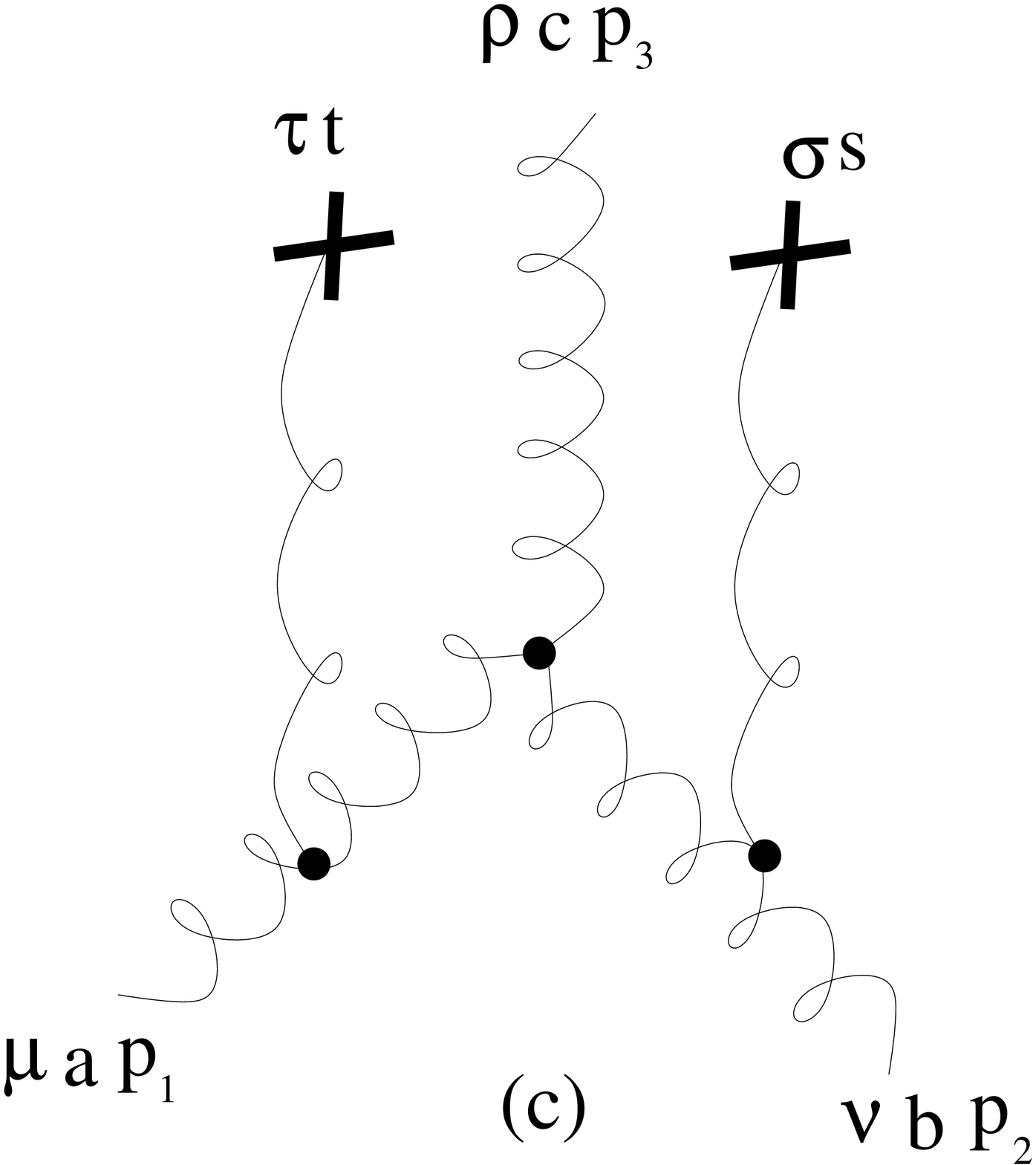} & &
\epsfxsize4.0cm\epsffile{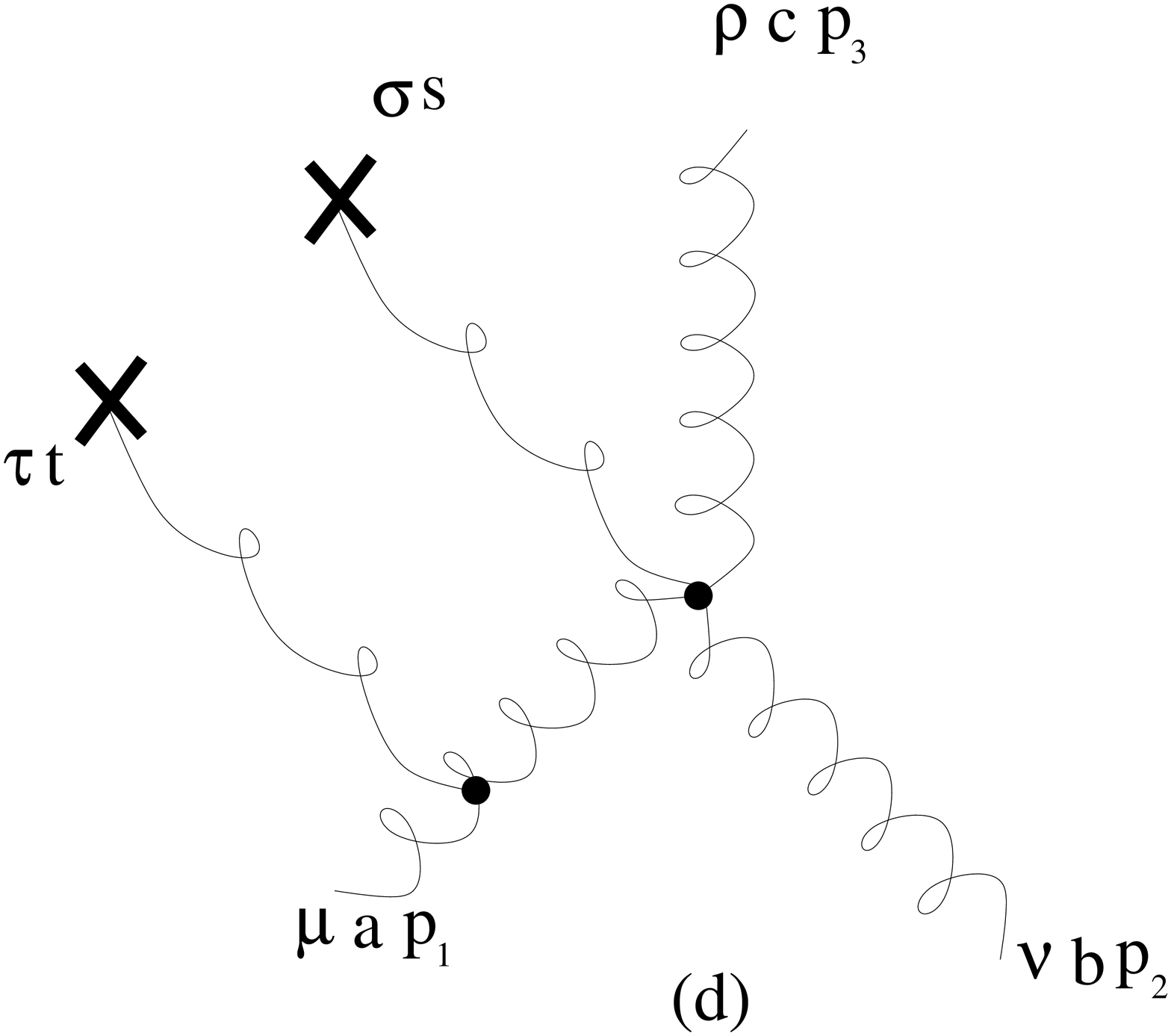} \\
\end{tabular}
\caption{\small {\it Five gluons diagrams to be computed, among all the 
possible new ones obtained by permutations of leg labels, in order to determine 
the tree-level OPE correction to the gluon three-point Green function.}}
\label{Fig2}
\end{center}
\end{figure}

\noindent Then, we follow from Eq. (\ref{OPEfield2}) a procedure to  obtain the
OPE perturbative coefficient $d_2$  analogous to that to the one previously used
for the gluon propagator. The diagrams to  compute are those plotted in Fig.
\ref{Fig2}. In total analogy to the gluon  propagator case, the result for the
tree-level OPE correction of the  three-point Green function, translated to
Euclidean metrics, leads to 

\beq
G^{(3)}(k^2,k^2,k^2)\ = \ G^{(3)}_{\rm pert}(k^2,k^2,k^2) \ + \ 
\frac{9g^3\langle A^2 \rangle}{4(N_c^2-1)} \ \frac{1}{k^2}  \; ;
\label{OPEescalar3p}
\eeq

\noindent where, again, $G^{(3)}_{\rm pert}$ stands for the purely perturbative 
three-point form factor. 

Then, we insert Eqs. (\ref{OPEescalar2p},\ref{OPEescalar3p}) in the MOM  coupling
constant definition, Eq. (\ref{gdef}), where the Green functions scalar form
factors  are automatically renormalized in MOM prescription at the scale $k^2$ by
the $3/2$-power  of $Z^{MOM}$, and finally obtain\footnote{The 
$O(\alpha)$-corrections in the second r.h.s. of Eq. (\ref{OPEalpha}) are of the
order  of the assumed theoretical uncertainty in our tree-level OPE correction.}

\beq
\as(k^2) = \left(\as(k^2)\right)_{\rm pert} \left[ 1 \ + \ 
\frac{g_R^2 \langle A^2 \rangle_R}{4(N_c^2-1)} \
\frac{1}{k^2} \left( 9 \ + O(\alpha) \ \right) \ \right] \; .
\label{OPEalpha}
\eeq

\noindent One interesting remark is that all the OPE contribution to $\as(k^2)$
defined by Eq. (\ref{gdef}) is coming from the explicit renormalization factor
$(Z^{MOM})^{3/2}$. After amputation of the external legs, diagrams (a) and
(b) in fig \ref{Fig2} are obviously canceled, and a direct cancellation of the
irreducible  diagrams (c) and (d) happens.

As already mentioned only\footnote{the two first coefficients, $\beta_0$ and
$\beta_1$ are  universal} the two-loop coefficient of the beta function is known
in  the symmetric MOM scheme. The gluon propagator anomalous dimension is known
at three loops and previous works \cite{propag} has revealed the important role
played by the third loop in its appropriate perturbative description.   Moreover,
a previous analysis \cite{poweral} including power corrections of $\astil(k)$, in
$\widetilde{\rm MOM}$ scheme, also stressed the necessity of working at
least at three-loops to get a good estimate of $\Lams$. This is why we shall add
to the two loop test  of OPE a three-loop one with a {\it fitted} $\beta_2$.  The
latter test is performed consistently at the three-loop level both for the gluon
propagator  renormalization constant, $Z^{MOM}(k^2)$, and for the perturbative
part  of $\as(k^2)$, which we will take to be the numerical inversion of

\beq
\Lams= \frac{k}{3.334} \exp\left (\frac{-2 \pi}{\beta_0
	      \as}\right)
\left[	      
  \left(\frac{\beta_0  \as}{4 \pi}\right)^2	      
  \left(1+\frac {\beta_1\as}{2\pi\beta_0}+
   \frac{\beta_2\as^2}{32\pi^2\beta_0}\right)
\right]^{\frac{\beta_1}{2\beta_0^2}}  
\nonumber \\
\times \exp
\left\{\frac{\beta_0\beta_2-4\beta_1^2}{2\beta_0^2\sqrt{\Delta}}\left[
\arctan\left(\frac{\sqrt{\Delta}}{2\beta_1+\beta_2\as/4\pi}\right)
-\arctan\left(\frac{\sqrt{\Delta}}{2\beta_1}\right)\right]\right\} \; \; ,
\nonumber \\
 \label{alpert}
\eeq

\noindent where $\Delta\equiv 2\beta_0\beta_2-4\beta_1^2$, the unknown 
coefficient $\beta_2$ being fitted. The interplay between the ignorance of
non-tree-level OPE correction and this fitted three-loop information is hoped
to be controlled by studying the stability of the fits for different momentum
windows.

The OPE approach used for propagator and symmetric three-point Green function is based on
the {\it factorization} of the non-perturbative soft gluon contribution (related
to the vacuum properties) and the perturbative hard gluons in the external legs
of the Green functions. In the $\widetilde{\rm MOM}$ case, things are a little different. 
That has to be adapted since one of the external  gluons
of the Green function carries  zero momentum. OPE large momentum expansion  can
nevertheless be done for the two other gluon fields carrying non-null momentum 
by following Eq. (\ref{OPEfield1}). As only an odd number of gluon field 
operators permits to build a Lorentz invariant {\it v.e.v.}, non-null 
contributions will come now for terms of the Wilson expansion with an even number
of  local operators.  Then, we will have

\beq
\langle T\left( 
\Am{\mu}{a}{-p} \ \Am{\nu}{b}{p} \Am{\rho}{c}{0} \right) \rangle \ = \
(\widetilde{c}_1)^{a b \rho'}_{\mu \nu c'}(p^2) \ \langle :\A{\rho'}{c'}{0}: 
\Am{\rho}{c}{0} \rangle 
+ \dots \; \; . 
\label{OPEpropagAS}
\eeq

\noindent As for the propagator or the symmetric vertex, it is immediate to prove
that $(\widetilde{c}_1)^{a b \rho'}_{\mu \nu c'}(p^2)$ takes the tree-level
expression for the  asymmetric three-point Green function (with the explicit
amputation of the zero-momentum gluon leg). Concerning subleading OPE corrections,
they should be obtained from local operators of higher dimensions, and then a ``{\it
zoo}'' of different condensates appear. In principle,  $\langle \ :A_{\mu'}^{a'}
A_{\nu'}^{b'} A_{\rho'}^{c'}: \widetilde{A}_\rho^c \rangle$,  $\langle \ :
\partial_{\mu'} \bar{c}^{a'} c^{b'} : \widetilde{A}_\rho^c \rangle$ or $\langle \
:\partial_{\rho'} \A{\mu'}{a'}{0} \A{\nu'}{b'}{0} :  \widetilde{A}_\rho^c \rangle$,
are all candidates to give non-null {\it v.e.v.} contributions. Furthermore,  the
problem due to the soft gluon in the vertex definition re-appears now  in a new way:
no {\it v.e.v} in the r.h.s of Eq. (\ref{OPEpropagAS})  involves directly the gluon
condensate $\langle A^2 \rangle$. Of course,  this does not forbid an explanation  of
power corrections in terms  of OPE for asymmetric $\astil$, but their
non-perturbative condensates  cannot be directly put  in relation with those from
symmetric $\as$ and gluon propagator. Consequently, with no additional
phenomenological assumption no consistent test shall come from the running coupling
constant in the asymmetric $\widetilde{\rm MOM}$ scheme and it will be left aside in
our present analysis.

\section{Fitting data to our ans\"atze} 
\label{sec:lattice}

Our goal is now to try a consistent description of lattice data for two- and
three-point Green functions introduced in ref.  \cite{frenchalpha,propag,poweral}
from the previously developed formulas for  power corrections. Details of the
lattice simulations, of the elaborated procedures to obtain an artifact-safe data
set or of the definition of regularization-independent objects permitting lattice
regularized  data to be matched to continuum quantities in any scheme, can be
found in those references. We directly show here the results of the matchings.

In a previous work \cite{poweral}, a fit of both the empiric coefficient for a 
power contribution and $\Lams$ were simultaneously performed by matching lattice
data  for $\astil(k^2)$ to a formula including non-perturbative additive power
corrections.  The $\Lams$  parameter thus obtained, $\Lams = 237^{+\ 3}_{-13}$ MeV,
is in total agreement with the one estimated by ALPHA group, $\Lams = 238(19)$
MeV.  The same kind of fit of $\astil(k^2)$ can be done by assuming the  power
correction to be proportional to $\astil(k^2)$ analogously to what is done in the
MOM case in eq.  (\ref{OPEalpha}). The fitted $\Lams$ parameter raises in the 
latter fit to  $\sim 250$ MeV. This  difference coming from the assumed
$\propto \astil(k^2)$ dependence of the $1/k^2$ term
 tells us something about a certain additional systematic
uncertainty, not taken into account in ref. \cite{poweral}, due to neglecting 
anomalous dimensions of power corrections. 

We will take the central value of the estimated  $\Lams$ in
ref. \cite{poweral}, $\Lams = 237^{+\ 3}_{-13}$ MeV, for 
$\alpha_s^{\rm MOM}$ and gluon  propagator.

\paragraph{Two loop fits}

We have fitted the gluon propagator and the symmetric coupling constant according 
to eqs. (\ref{OPEescalarR2p},\ref{ctree}) and (\ref{OPEalpha}) with the leading
perturbative parts computed to two loops. In both cases we fit the condensate.
The result is:
\beq
{\rm propagator :}\quad g^2_R\langle A^2 \rangle_R = (2.32(6) {\rm GeV})^2;\qquad
{\alpha^{\rm MOM} :}
\label{restwoloop}\quad g^2_R\langle A^2 \rangle_R = (4.36(12) {\rm GeV})^2.
\eeq 
with $\chi^2/d.o.f \simeq 1.2$. 
The clear disagreement between both estimates could be a sign that we are 
in the situation indicated in \cite{MartiSach} that the perturbative order
is too low to give an acceptable precision in the estimate of the power
corrections. We need to check if going to three loops is enough.

\paragraph{Three loops fit} $\beta_2^{\rm MOM}$ is not known. Then, our strategy will
be to extract the value for the gluon condensate, $\langle A^2 \rangle$, from the
matching of (\ref{OPEescalarR2p},\ref{ctree}) to gluon propagator lattice data; and
then, by using that value, to search whether any sensible value of the unknown MOM
coefficient $\beta_2$   allows to reproduce data for $\as(p)$.  

\begin{figure}[hbt]
\hspace*{-1.3cm}
\begin{center}
\begin{tabular}{ll}
\epsfxsize6.5cm\epsffile{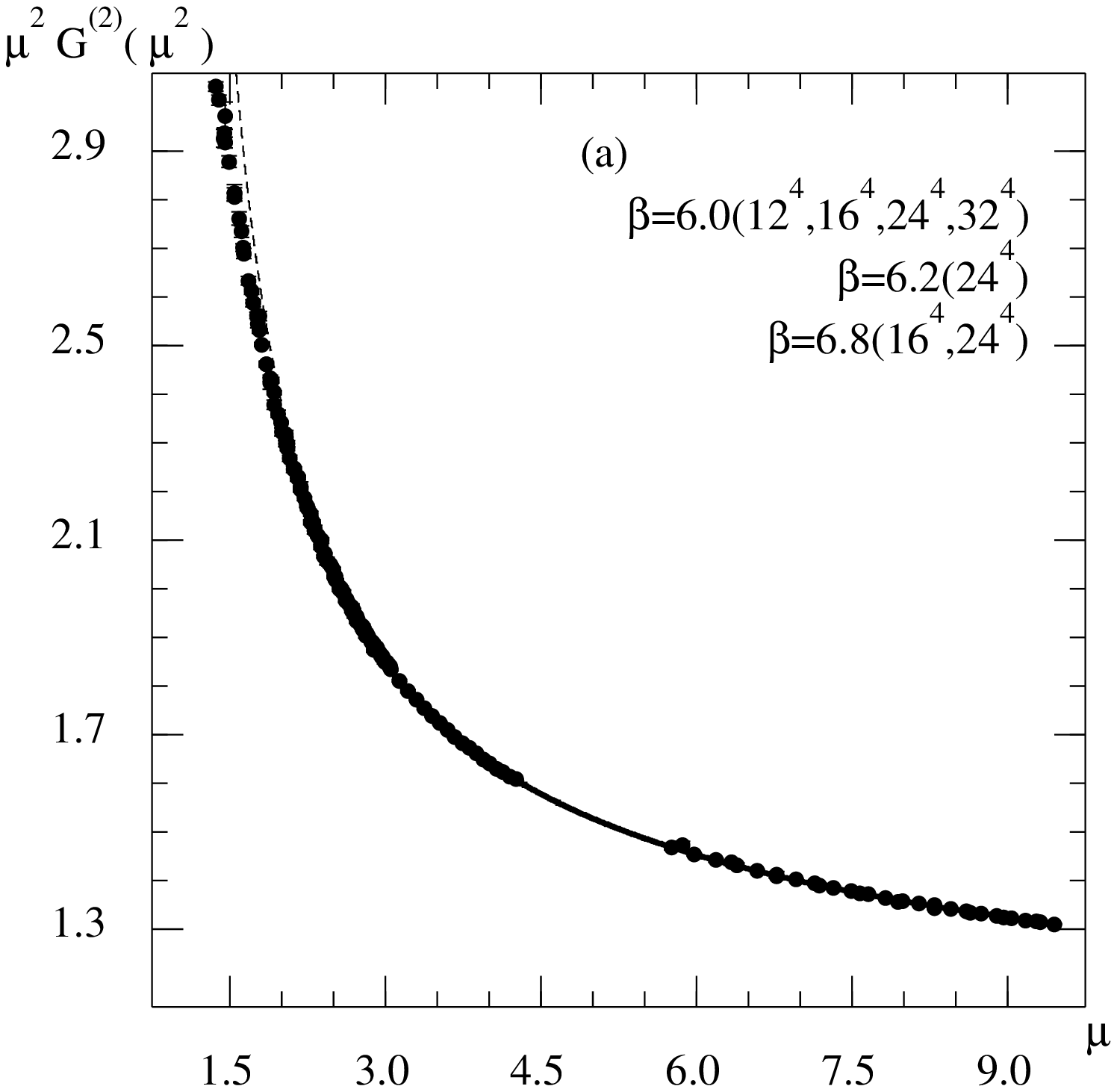} & 
\epsfxsize6.5cm\epsffile{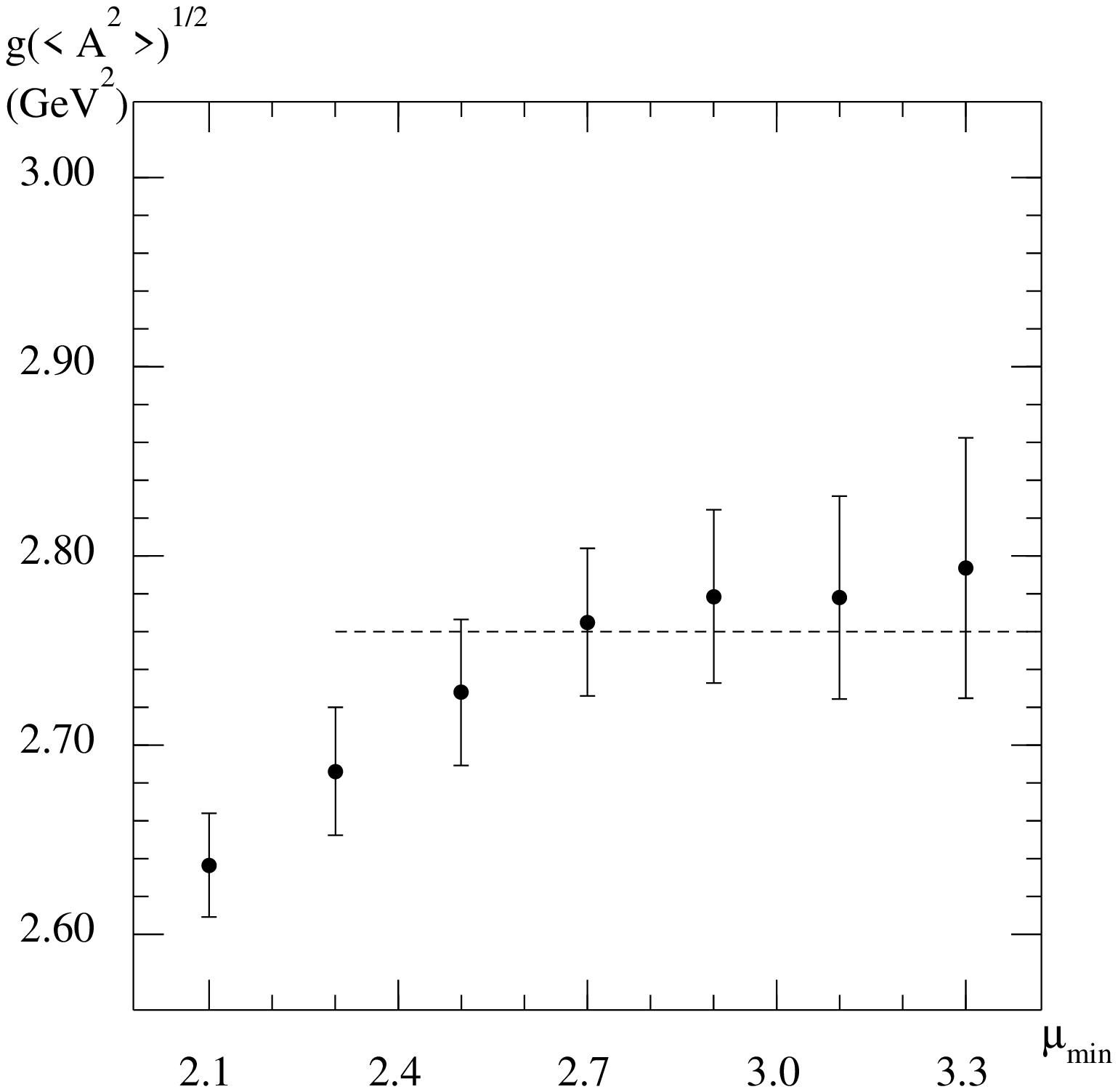} \\
\end{tabular}
\caption{\small {\it Formula (\ref{OPEpropagfin}), with the assumed value of
$\Lams$  ($237$ MeV), fits impressively the data for {$g^2 \langle A^2 \rangle =
(2.76(4) GeV)^2$} over a large momenta window from $2.6$ GeV to $\sim 10$ GeV
(fig. a). The optimal window is fixed by requiring a reasonable stability of the
fitted parameter ({$g^2 \langle A^2 \rangle$}) for  different choices of the
lower momentum (fig. b).}}
\label{Fig4}
\end{center}
\end{figure}

In fig. \ref{Fig4} we plot the best fit of eq. (\ref{OPEescalarR2p},\ref{ctree}), for  
\beq g^2_R
\langle A^2 \rangle_R \ = (2.76(4) \rm{GeV})^2,  
\eeq 
to gluon propagator data.
This error is computed by  jacknife's method. The $\chi^2$ is $0.86$ {\it per
d.o.f.} in that fit, slightly better than for the two-loop fit.
It is amusing that the three-loop condensate stemming from the propagator
is not so different from the two-loop one.

\begin{figure}[hbt]
\hspace*{-1.3cm}
\begin{center}
\begin{tabular}{ll}
\epsfxsize6.5cm\epsffile{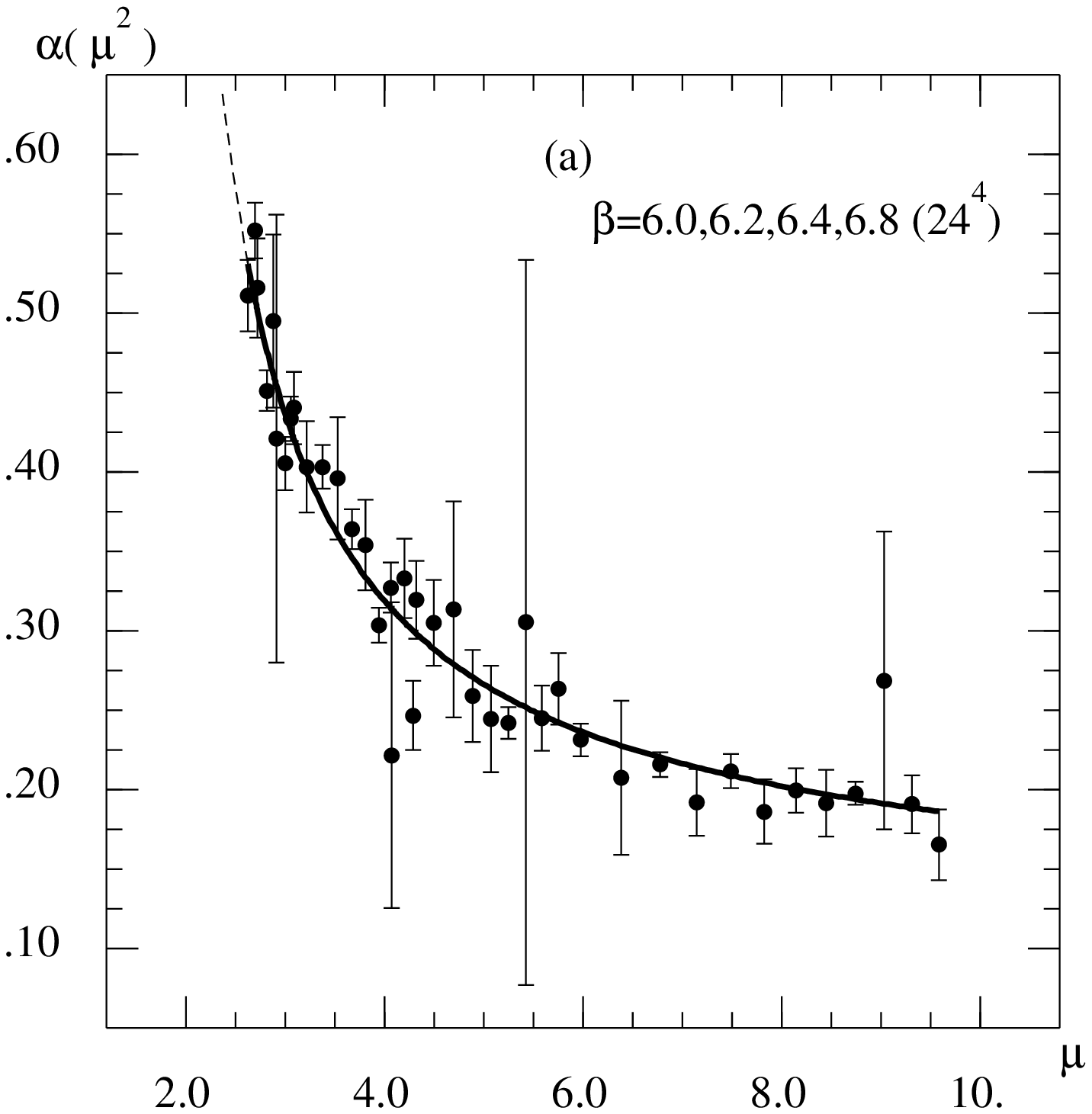} & 
\epsfxsize6.5cm\epsffile{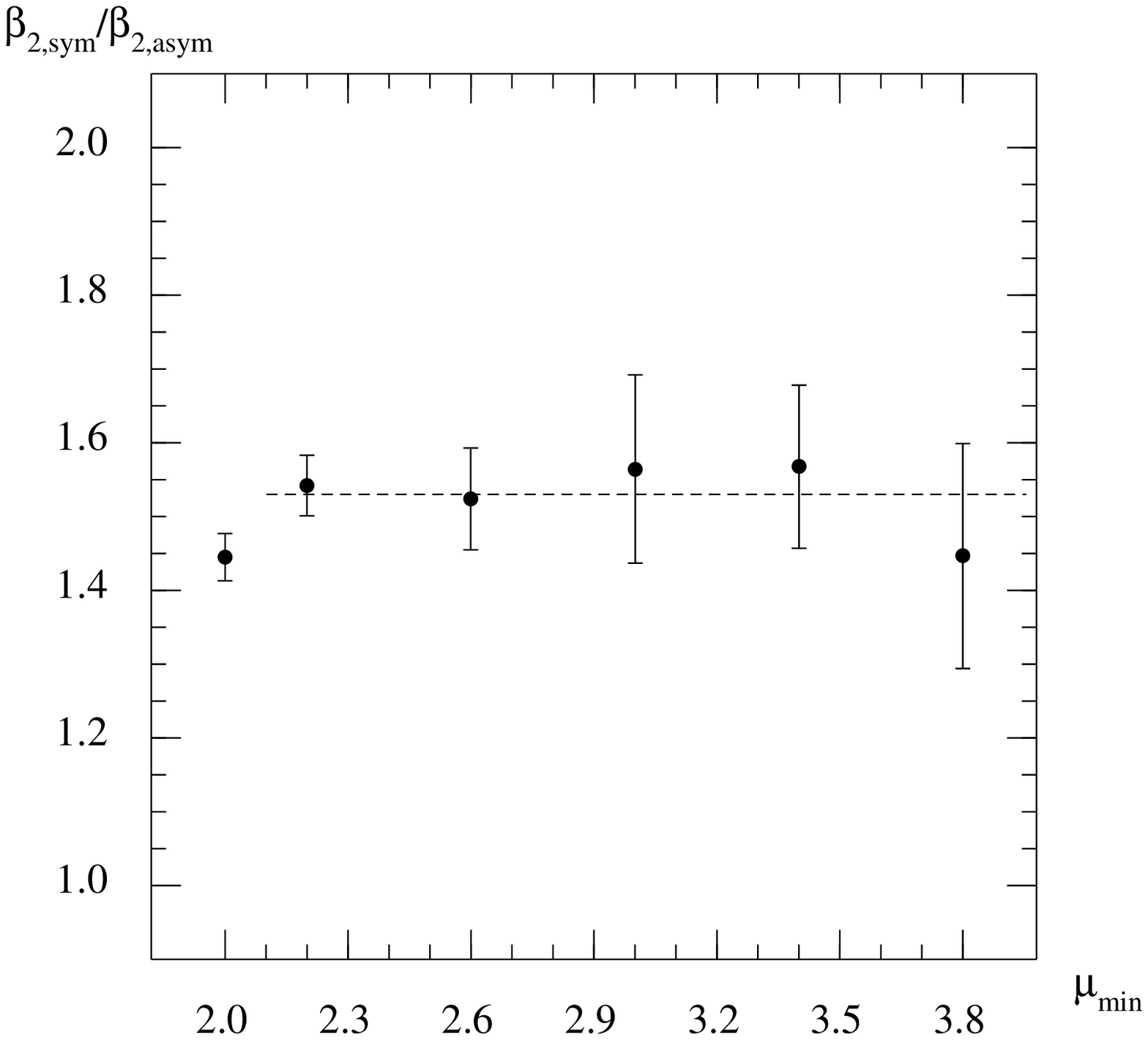} \\
\end{tabular}
\caption{\small {\it $\Lams$ assumed to be $237$ MeV and $g^2 \langle A^2 \rangle = 0.236(7)$
from gluon propagator analysis, the best fit of Eq. (\ref{OPEalpha}) to lattice data (gives 
$\beta_2^{\rm MOM}= 1.53 \times \beta_2^{\widetilde{\rm MOM}}$ (fig. a). Fig. b shows the
stability of this parameter faced to different choices of the lower momentum of the 
fitting window.}}
\label{Fig5}
\end{center}
\end{figure}

With this  value of the gluon condensate we find that
\beq
\beta_2^{\rm MOM} = 1.53(7) \times \beta_2^{\widetilde{\rm MOM}}
\sim 7400 \pm 300
\label{resbeta2}
\eeq 
 optimises the quality matching 
($\chi^2/{\it d.o.f.}=1.1$) of formula (\ref{OPEalpha}) to lattice data 
(see fig. \ref{Fig5}).

\section{Discussion and conclusions}
\protect\label{sec:conclusions}

A certain underestimation of the errors presented in the previous section should
be admitted as a  consequence of fixing $\Lams$ to its central value 237 MeV  of
\cite{poweral} without errors. We can correct for this by the same fitting
strategy applied with different starting assumptions for  $\Lams$, within its
error interval. this results in a a wider dispersion:
 \beq
  \beta_2^{\rm MOM}/\beta_2^{\widetilde{\rm
MOM}}= 1.5(3).
\eeq
  $\Lams$ and $\beta_2^{\rm MOM}$ appear to be strongly correlated. The goal of
the present paper is less to give a reliable prediction of a certain perturbative
parameter, as $\beta_2^{\rm MOM}$, than to carry out a preliminary study of the
inner consistency from a description in terms of OPE predictions of  lattice
data. Let us repeat, the knowledge of three-loop information in the MOM  scheme
and the anomalous dimensions of the power corrections are prerequisite to any
firm conclusion\footnote{A very recent lattice calculation within the 
two-dimensional non-linear sigma model \cite{CMP}
insists on the necessity of knowing the OPE coefficients at least to one loop
to reach agreement with OPE predictions.}. The comparison of our estimate of $\beta_2^{\rm MOM}$ and the
future perturbative computation of this parameter\footnote{A computation of
$\beta_2$ may appear soon \cite{chet}.} are totally necessary to conclude whether
signs of such a consistency are present or not.

Our analysis of lattice data from gluon propagator and running coupling constant
unequivocally  proves power corrections ($1/p^2$) to be necessary to match the
data. The dimension $2$ of power corrections clearly manifests the presence of
non-gauge-invariant condensates  (in particular, in our pure gauge QCD analysis,
$A_\mu A^\mu$), if OPE is invoked to explain the results.  Since we work in the
Landau gauge the appearance of such a non-gauge-invariant condensate is in full
agreement with all theoretical expectations.  The value for $\sqrt{g_R<A^2>_R}$ 
around 3 GeV might seem unexpectedly large but this is a subjective matter.   The
perturbative Wilson coefficients for the OPE of the involved Green functions have
been theoretically computed. Then, in order to satisfy the universality of the 
non-perturbative gluon condensate appearing directly in OPE of both gluon
propagator and MOM coupling constant, we need to assume  $\beta_2^{\rm
MOM}=7400(1500)$. The perturbative computation of this parameter  as well as that
of the anomalous dimensions of the power corrections will either  exclude the
interpretation of power corrections just in terms of universal gluon condensates
at three loops, or on the contrary throw a positive sign supporting the usual
``{\it QCD sum rules}'' procedure to connect non-perturbative effects for
different  correlation functions. All the theoretical tools for such a OPE 
testing procedure have been developed in the present paper.

\section{Acknowledgments} 
We specially thank D. Becirevic for thorough discussions at the earlier stages of the 
work. We are also indebted to Y. Dokshitzer and G. Korchemsky 
for several inspiring comments. J. R-Q is indebted to Spanish 
Fundaci\'on Ram\'on Areces for financial support.
These calculations were performed on the QUADRICS QH1 located 
in the Centre de Ressources
 Informatiques (Paris-sud, Orsay) and purchased thanks to a
  funding from the Minist\`ere de
  l'Education Nationale and the CNRS.

\end{document}